\documentclass[a4paper]{jpconf}
\usepackage{graphicx}
\usepackage[latin1]{inputenc}
\usepackage{multirow}

\renewcommand{\AM}{$A\,$MeV}

\newcommand{\sys}[4]{\ensuremath{^{#1}}#2+\ensuremath{^{#3}}#4}

\begin{document}
\title{Status and performances of the FAZIA project}

\author{M F Rivet$^1$, E Bonnet$^2$, B Borderie$^1$, R Bougault$^3$, P Edelbruck$^1$,
N~Le~Neindre$^3$, S Barlini$^4$, S Carboni$^4$, G Casini$^4$, E Rosato$^5$, 
G~Tortone$^5$  for the FAZIA Collaboration}

\address{$^1$ Institut de Physique Nucl\'eaire, CNRS-IN2P3 and University
Paris-Sud 11, F-91406 Orsay France}
\address{$^2$ GANIL, CEA-DSM/CNRS-IN2P3, F-14076 Caen Cedex, France}
\address{$^3$ Laboratoire de Physique Corpusculaire, CNRS-IN2P3, ENSICAEN
and University of Caen  F-14050 Caen France}
\address{$^4$ Universit\`a di Firenze and INFN Sez. Firenze, I-50019 Sesto
Fiorentino,  Firenze, Italy}
\address{$^5$ Dipartimento di Scienze Fisiche e Sezione INFN, Universit\`a
di Napoli ''Federico II'', I-80126 Napoli, Italy}

\ead{rivet@ipno.in2p3.fr}

\begin{abstract}
FAZIA is designed for detailed studies of the isospin degree of freedom,
extending to the limits the isotopic identification of charged products from
nuclear collisions when using silicon detectors and CsI(Tl) scintillators. We
show that the FAZIA telescopes give isotopic identification up to Z$\sim$25
with a $\Delta$E-E technique. Digital Pulse Shape Analysis makes possible
elemental identification up to Z=55 and isotopic identification for Z=1-10
when using the response of a single silicon detector. The project is now in
the phase of building a demonstrator comprising about 200 telescopes.
\end{abstract}

\section{Introduction}
With the advent of exotic beams delivered by present and future accelerator
facilities, it is a challenge for physicists to design detection arrays able
to completely identify (atomic number and mass, energy) the nuclear reaction 
products, taking thus full advantage of these new beams. 
Indeed one of the present motivations for studying nuclear collisions is 
to improve the knowledge of the nuclear equation of state, and particularly 
of the density dependence of its symmetry energy term, 
E$_{sym}$~\cite{Bar05,Bao08,Riv09}. In this aim, 
at moderate and low energies, dissipative collisions induced by exotic
projectiles should provide new information on the neck-instabilities, 
and on the achievement of isospin equilibration, both directly depending on 
E$_{sym}$.
In the same energy domain a study of the de-excitation properties of 
isotopic chains of compound nuclei, with complete isotopic identification of
the emitted nuclei and of the final residue~\cite{MarCNR09}, will allow to
test the N/Z dependence of the level density parameter; this will give
unique information on the temperature dependence of E$_{sym}$. Pushing CN
towards the drip lines will permit to explore decay modes that cannot be
predicted by the Weisskopf theory (clustering, multifragmentation) and thus
to test the validity domain of the basic assumptions of that theory.
With an increased knowledge of the de-excitation of hot nuclei, when going
around the Fermi energy, thermodynamical 
properties of multifragmenting nuclear systems (excitation energy, caloric curves) 
should be more accurately determined~\cite{Viol06,Bor08}, providing more
precise values of limiting temperatures. Measurements of
fragment isotopic distributions should also permit to constrain  E$_{sym}$.

At present time, the most powerful 4$\pi$ arrays suffer from several
drawbacks. INDRA~\cite{I3-Pou95}, is composed of Ionisation Chamber-Silicon-CsI(Tl)
telescopes and uses $\Delta$E-E and light shape analysis to identify nuclei.
It well identifies the atomic number of the charged products up
to Z$\sim$60, with a threshold of $\sim$ 1~\AM{},
but its isotopic identification is limited to Z=6. 
CHIMERA~\cite{R5-Pag04} is an ensemble of Silicon-CsI(Tl) telescopes which
identifies nuclei through time of flight, $\Delta$E-E and light shape analysis 
techniques. It has very low thresholds for mass identification, but much higher 
ones for Z-identification. CHIMERA gives isotopic identification, Z and A,  up to
Z=13~\cite{Lom11}.
Even specific correlators such as LASSA~\cite{Dav01}, based on $\Delta$E-E
identification, have an isotopic separation only up to Z=9. The reader is sent
to ref.~\cite{Sou06} for a more exhaustive detector review. The unsurpassed
 nucleus identifiers are still spectrometers, which provide mass
charge, energy and velocity of nuclei with a high resolution. They cover
however a very small solid angle and cannot fully analyse many-body
events, unless they are coupled to other detection arrays.

In this context a group of french and italian physicists studying nuclear
dynamics and thermodynamics started, at the dawn of the century, a vigourous
R\&D aiming at obtaining isotopic identification up to mass 50 over a large
solid angle. The main idea was to examine the detailed shape of the current
signal created in a silicon detector by the passage of a charged particle,
by sampling it with a fast ADC.
This group later became the FAZIA Collaboration, which nowadays comprises
about 60 physicists from France, Italie, Poland, Romania and Spain.

\section{The FAZIA telescope} \label{Tele}
FAZIA is intended to work for studying reactions from the Coulomb barrier 
up to 100~\AM{}. It was thought as a 4$\pi$ array comprising a large
number (several thousands) of three-stage telescopes, two silicon detectors 
(Si1: 300~$\mu$m and Si2: 500~$\mu$m) followed by  a CsI(Tl) scintillator readout 
by a photodiode. 
Gas detectors are avoided because of their mechanical size. The desired low
identification thresholds are provided by pulse shape analysis (PSA) of the
signals of the first silicon, with the help of time of flight measurement
for very slow particles~\cite{Paus94,Mut00}. 
In order to improve the performances of present
arrays, a lot of work was performed on the response of CsI(Tl) scintillators
and above all on that of silicon detectors.

\subsection{Silicon detectors: the lessons of R\&D} \label{Si}
When one wants to use the shape of the signal created by a charged particle
traversing a silicon detector of finite area, it is mandatory that the
detector response be independent of the impact position. This implies that
the field should be homogeneous in the bulk of the detector.
 An answer to this requirement is obtained by using n-type implanted silicon 
 doped  with phosphorus through neutron transmutation;
this process provides silicon wafers with resistivity around 3000~$\Omega$.cm
and resistivity fluctuations smaller than 4\% FWHM (against 30\% for standard
silicon wafers). Another constraint is that particles must not cross the
detector following a crystal axis or plane: as known for 50 years, detectors
must be cut off-axis to avoid channeling effects which induce large
fluctuations in the response to monoenergetic particles~\cite{Bard09,Bar11}.
Note that this constraint limits the angular aperture of a telescope; the
FAZIA telescopes have a square area 2$\times$2 cm$^2$. They will be mounted
between 1 and 1.2~m from the target.
Finally the bias voltage on the detector is kept constant by permanently 
monitoring the reverse current and correcting the applied voltage.

We chose a reverse mounting so that particles enter through the
low-field side of the detector, in order to maximize the shape differences
between elements and isotopes~\cite{Mut00}. However although in this last
reference overbiasing detectors was recommended, we found that a better
identification was obtained when using the detector depletion voltage, or
even slightly less. A compromise has still to be found between shape
discrimination and good timing for time of flight.

\subsection{The CsI(Tl) scintillators} \label{CsI}
The scintillator length was fixed to 10~cm, to make FAZIA operational up
to 100~\AM{} by stopping all light charged particles. 
Having such a length, the crystals are tapered to keep the same angular
aperture at both ends. In order to favor scintillation efficiency for heavy ions, 
we chose a Tl concentration of $\sim$2000 ppm~\cite{Deg08}. The light
response and resolution, as well as its homogeneity along the detector length
were improved by wrapping the crystals with a highly reflective material, 3M
Vikuity$^{TM}$ ESR~\cite{Bed04}. 
The CsI(Tl) are coupled to 2$\times$2 cm$^2$ photodiodes.

\subsection{Front-end electronics} \label{FEE}

\begin{figure} \centering
      \centering \includegraphics[width=0.9\textwidth]{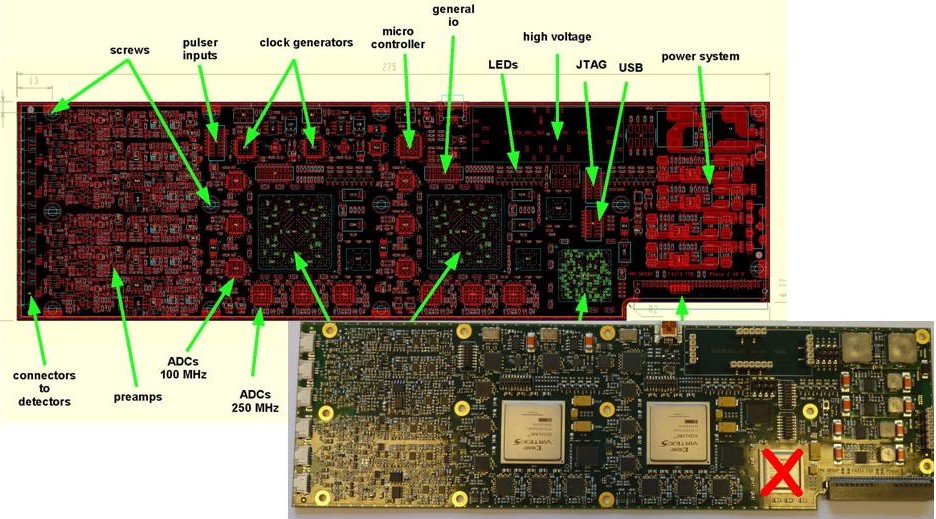}
      \caption{Pattern and photo of the prototype of the FEE board of FAZIA.
      The crossed small FPGA is removed in the final board.}
      \label{Fig:FEE}
\end{figure}  

The Front End Electronics board of FAZIA, designed at IPN Orsay, houses all the 
necessary modules for complete processing of the three detector signals of 
two telescopes.  The prototype version is represented
in figure~\ref{Fig:FEE}. The detectors are connected to the preamplifiers,
located on the board, by kapton strips. The preamplifiers are derived from
the PACI~\cite{Ham04}, without the current output; the current signal, I, is now
obtained in the FEE board by analogically deriving the charge signal, Q. The
gain on the charge output is 2~mV/MeV.  

\begin{table}
\caption{\label{table:FEE} Characteristics of the fast ADC's for processing the
different signals of the silicon detectors and of the photodiodes. E range
indicates the full scale energy range.}
\begin{tabular}{lccccc} \br
& \multicolumn{3}{c}{Q signal} & \multicolumn{2}{c}{I signal} \\ \mr
Detector & E range & sampling rate & resolution & sampling rate & resolution \\ \mr 
\multirow{2}{*}{Si1}& 250 MeV & 250 Ms/s & 14 bits   & \multirow{2}{*}{250 Ms/s} & 
\multirow{2}{*}{14 bits}   \\
                    & 4 GeV & 100 Ms/s & 14 bits   &       &    \\
Si2                 & 4 GeV & 100 Ms/s & 14 bits   & 250 Ms/s & 14 bits   \\
CsI(Tl) + PD        & 4 GeV & 100 Ms/s & 14 bits   &  &    \\ \br

\end{tabular}
\end{table}

Each signal is sent to a fast ADC (see table~\ref{table:FEE}), apart from the 
Q signal of Si1 which is
processed twice, with a high gain for light particles and fragments, and with
a low gain for heavier ions, in order to improve PSA and Time of Flight in both 
cases (see sect.~\ref{PID}). Thus a FEE board comprises 6$\times$2 14 bit ADC's.
The high voltages for the two silicon detectors are built up locally in the FEE
card and are individually controlled (section~\ref{Si}); the card performs
the voltage compensation due to reverse current. One FPGA is in charge of a
preprocessing of all signals for each telescope. It stores and reads out the Q and I
waveforms. It gives a zero-level trigger which
signals that the \emph{telescope} was hit. It then provides a precise value
of the energy deposited in Si1, Si2 and CsI(Tl)+PD by means of trapezoidal filters.
The FEE card also holds one pulser for linearity calibration purpose and clocks 
to synchronize the different signals.
All information from the FPGA is transmitted to the acquisition system
through a block card (section~\ref{Demo}), the connector between both cards
is visible on the low rhs of the photo.

\section{Performances} \label{Perf}

\subsection{Particle identification} \label{PID}

\begin{figure} \centering
   \begin{minipage}[c]{0.49\textwidth}
      \centering \includegraphics[width=1.25\textwidth]{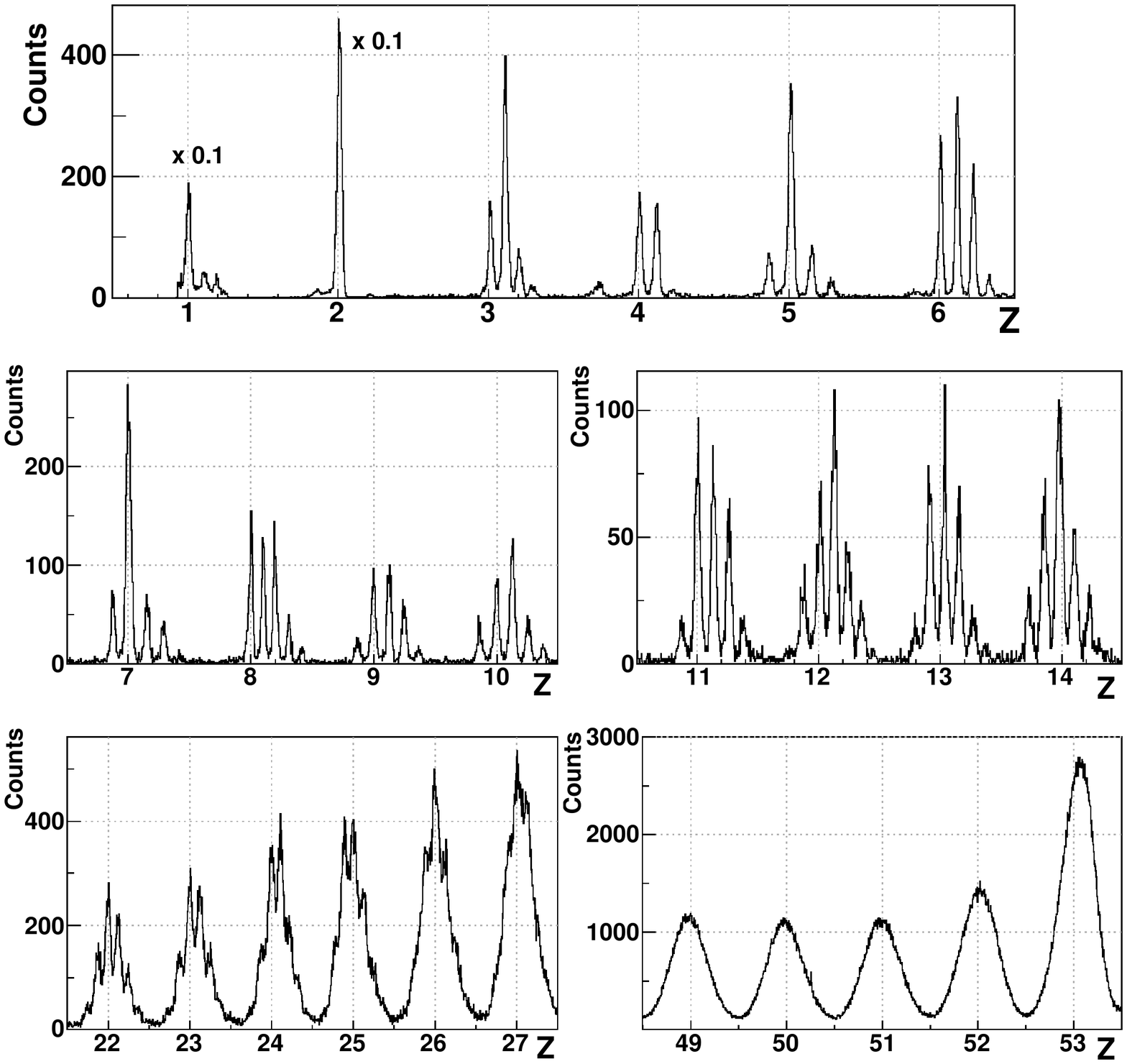}
      \caption{Distribution of particle identification function from
      a $\Delta$E-E map between the Si1 and Si2 signals. From~\cite{Car12}.}
      \label{Fig:P1}
   \end{minipage}%
    \hspace*{0.02\textwidth}
   \begin{minipage}[c]{0.49\textwidth}
      \centering \includegraphics[width=1.05\textwidth]{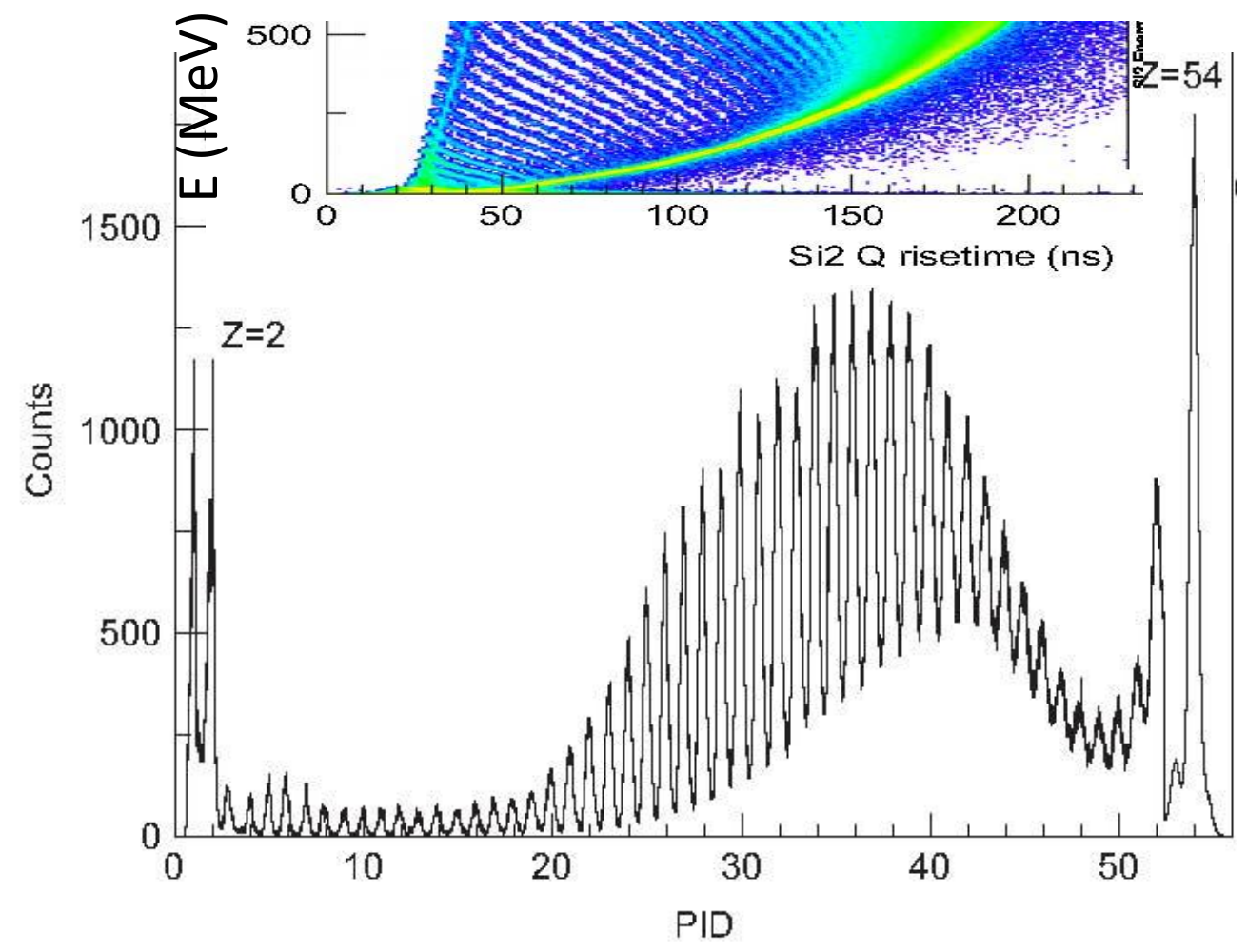}
      \caption{Particle identification obtained with the PSA method ``E vsI$_{max}$'' 
      in the second silicon alone. The inset shows the low
      energy part of the E vs $\tau _Q$ map for the same detector.
      Adapted from~\cite{Car12}.} \label{Fig:P2}
   \end{minipage}
\end{figure}

We found that improving the qualities of the silicon detectors
of the telescope also greatly enlarged the isotopic identification when using
the $\Delta$E-E method for Si1 vs Si2 responses.
Figure~\ref{Fig:P1} shows that masses are separated for Z up to 23 with this
technique. We have found isotopic identification on the same mass range when
using the $\Delta$E-E method with (Si1+Si2) vs CsI(Tl)~\cite{Car12}.

\begin{figure}
   \begin{minipage}[t]{0.49\textwidth}
      \centering \includegraphics[width=1.1\textwidth]{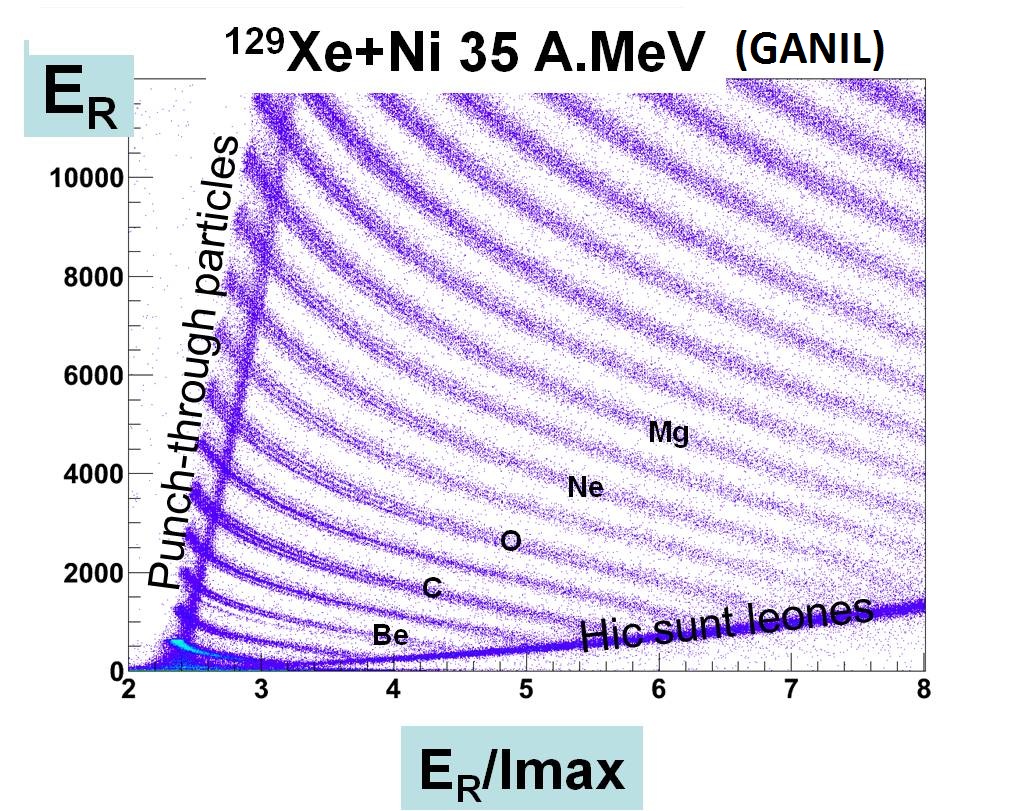}
      \caption{PSA: E vs E/I$_{max}$ for particles produced in a 35~\AM{} 
      \sys{129}{Xe}{58}{Ni}. The full energy range of the ordinate axis is 
      700 MeV. } \label{Fig:P3}
   \end{minipage}%
    \hspace*{0.02\textwidth}
   \begin{minipage}[t]{0.49\textwidth}
      \centering \includegraphics[width=1.05\textwidth]{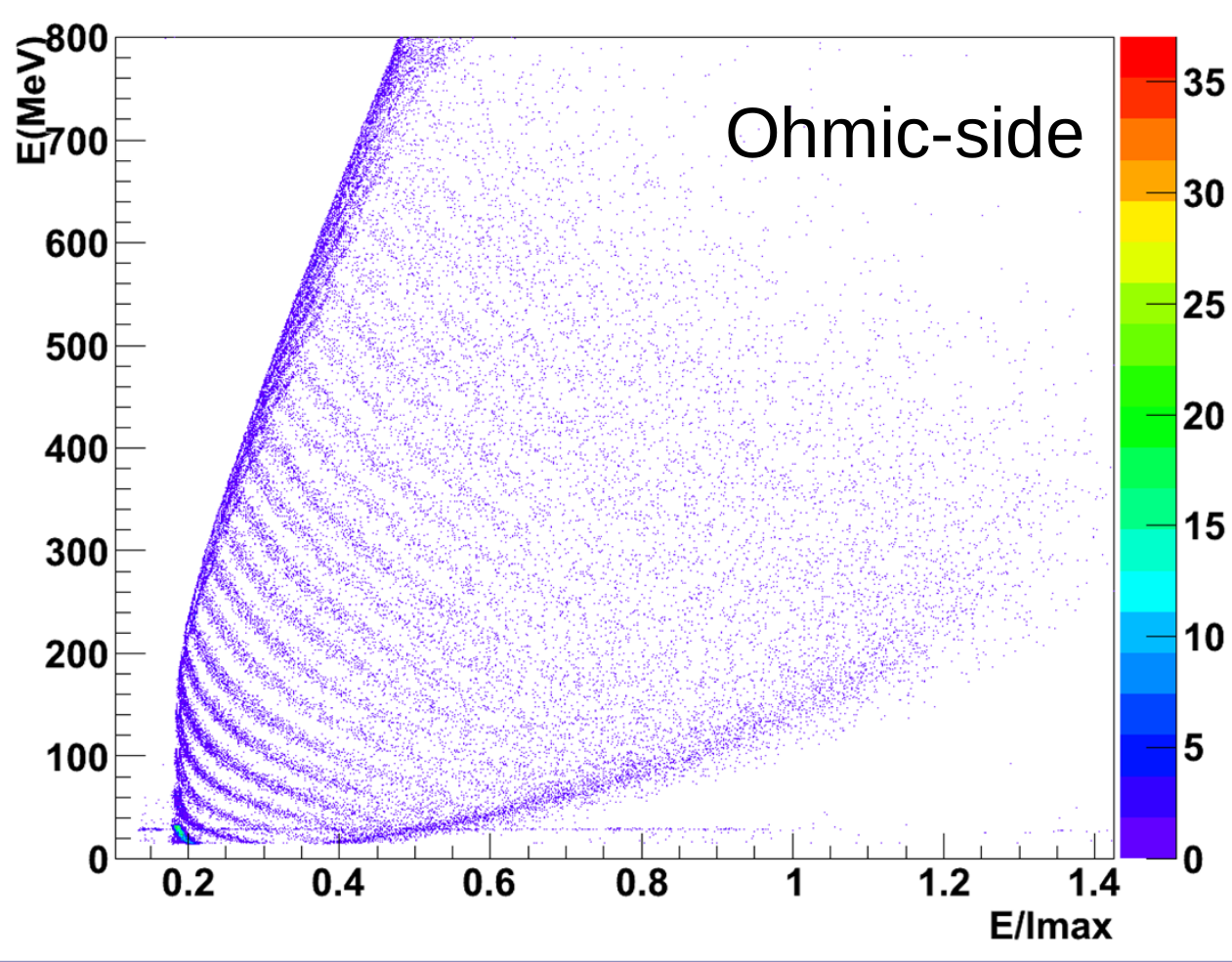}
      \caption{PSA: E vs E/I$_{max}$ for particles produced in a 35~\AM{} 
      \sys{129}{Xe}{58}{Ni} and detected in a DSSSD.} \label{Fig:P4}
   \end{minipage}%
\end{figure}

Identifying nuclei stopped in one silicon implies using two parameters
provided by this single detector. One of these is the measured energy of
a particle; for the second we chose either the rise time of the charge signal,
$\tau _Q$, or the maximum amplitude of the current signals (I$_{max}$
or E/I$_{max}$).
The Particle IDentification (PID) data presented in figure~\ref{Fig:P2} 
were obtained with a detector for
which the latter method was more performant, and Z identification
is observed up to Z=54~\cite{Car12}; in that case we performed PSA on the second 
silicon, in which most of the produced nuclei were stopped. 
Indeed figures~\ref{Fig:P1} and \ref{Fig:P2} were obtained with the same
telescope detecting nuclei produced in  35~\AM{} \sys{129}{Xe}{124}{Sn} 
reactions. The 306$\mu$m second silicon\footnote{this telescope comprised 
two 300~$\mu$m silicon, it was thus not a ``standard'' FAZIA telescope} 
was exceptionnally good, having a resistivity $\rho$ = 3000~$\Omega$.cm with 
a FWHM of the $\rho$ distribution across the detector of
0.6\%~\cite{Bar09}; it was biased 3\% above the depletion voltage. 
Figure~\ref{Fig:P2} also shows that no mass identification is visible, even
for the lightest elements. This is in disagreement with some of our  
results, where isotopes were seen for Z up to Z=7-8, see figure~\ref{Fig:P3}
and ref~\cite{Bar11}. This discrepancy was attributed to the gain of the
preamplifier, which was 3 times larger for the data of figure~\ref{Fig:P3} than 
for those displayed in figure~\ref{Fig:P2}. This led us to adopt two gains for 
the coding of the energy signal of the first silicon detector of each FAZIA 
telecope.

\subsection{Foreseen possible improvements}
In view of possible developments aiming at an increased granularity, we have
tested the PSA of the response of Double Sided Strip Silicon Detectors (DSSSD). 
The map displayed in figure~\ref{Fig:P4} was obtained with a 500~$\mu$m nTD
silicon, with 16 strips on each side. Particles entered through the
low-field (ohmic) side. Same results were obtained when taking signals
from the P-side or the ohmic-side. The figure shows that elements are 
identified with resolution and range similar to those of the other detectors.

Another solution which is thought of, because if would save quite an amount
of money for the 4$\pi$ array, is to replace the FAZIA modules at backward
angles (past 90$^o$ in the laboratory) by Single Chip Telescopes, SCT. This
arrangement suppresses one electronic channel, because the silicon detector
preceding the CsI(Tl) can be used both as a detector and as a photodiode for
the scintillator light readout. SCT's were tested during the FAZIA R\&D and,
using suitable algorithms, it was shown that performances comparable to those
of a FAZIA telescope were obtained for H and He isotope identification, 
whereas they are not as good for light fragments, particularly at low 
energy~\cite{Pas12}. For this reason SCT should be positioned at angles
where only few and light fragments are emitted.

\subsection{Energy identification thresholds} \label{Seuils}
We can observe in the inset of figure~\ref{Fig:P2}, showing part of a E vs $\tau _Q$ 
map, that at low energy nuclei can not be identified with a PSA method: all
ridge lines converge into a single locus. We quantified the energy threshold for
Z-identification, using a factor of merit, 
$FoM = (\langle PID_2 \rangle - \langle PID_1 \rangle) / (FWHM_1 + FWHM_2)$
where $\langle PID_i \rangle$ are the average PID values for nuclei Z$_1$ and
Z$_2$=Z$_1$+1, and the $FWHM_i$ are the full widths at half maximum of the 
PID peaks. Two elements are considered as well separated when $FoM >$0.7.

\begin{figure}
   \begin{minipage}[b]{0.5\textwidth}
      \includegraphics[width=\textwidth]{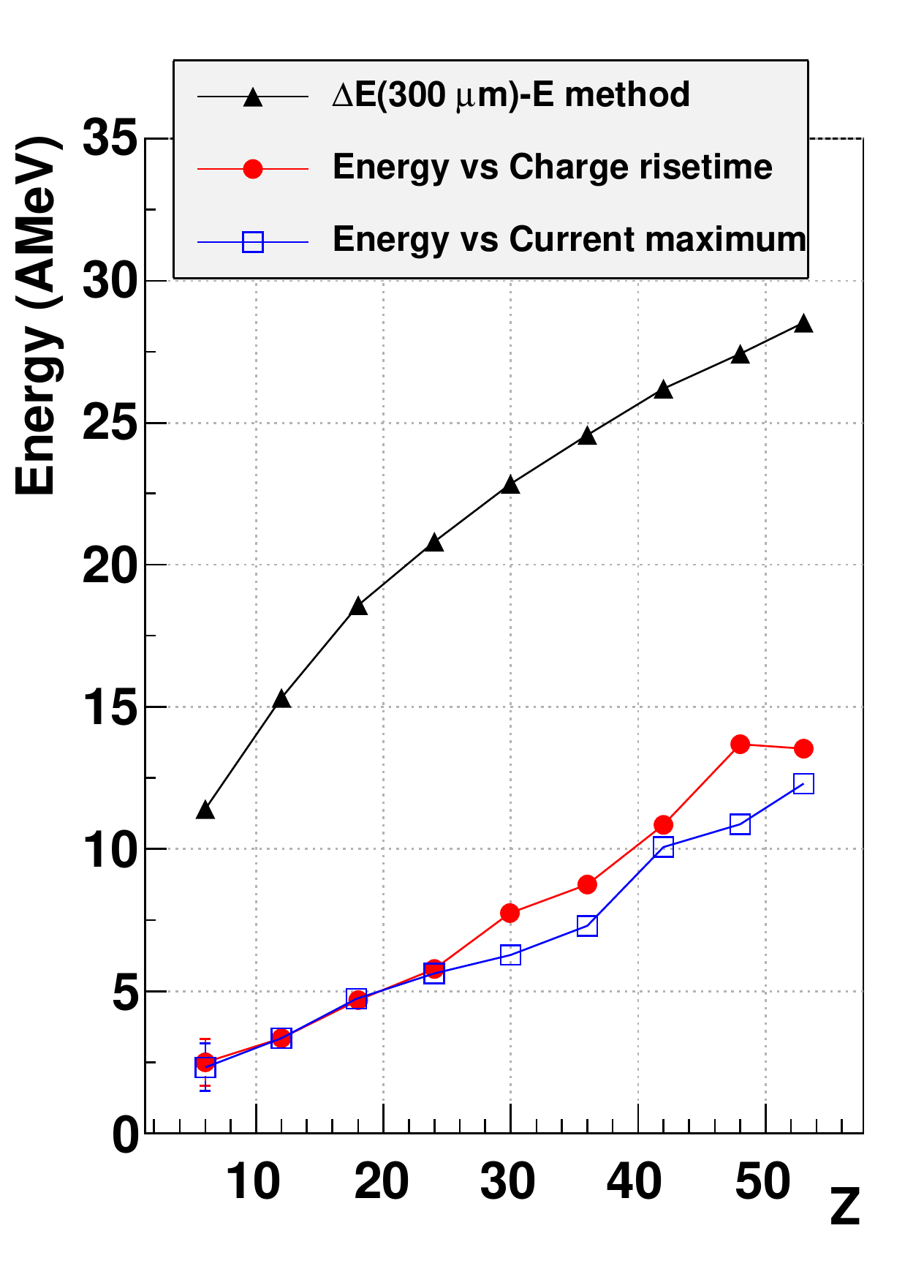}
   \end{minipage}%
   \begin{minipage}[b]{0.5\textwidth}
       \includegraphics[width=\textwidth]{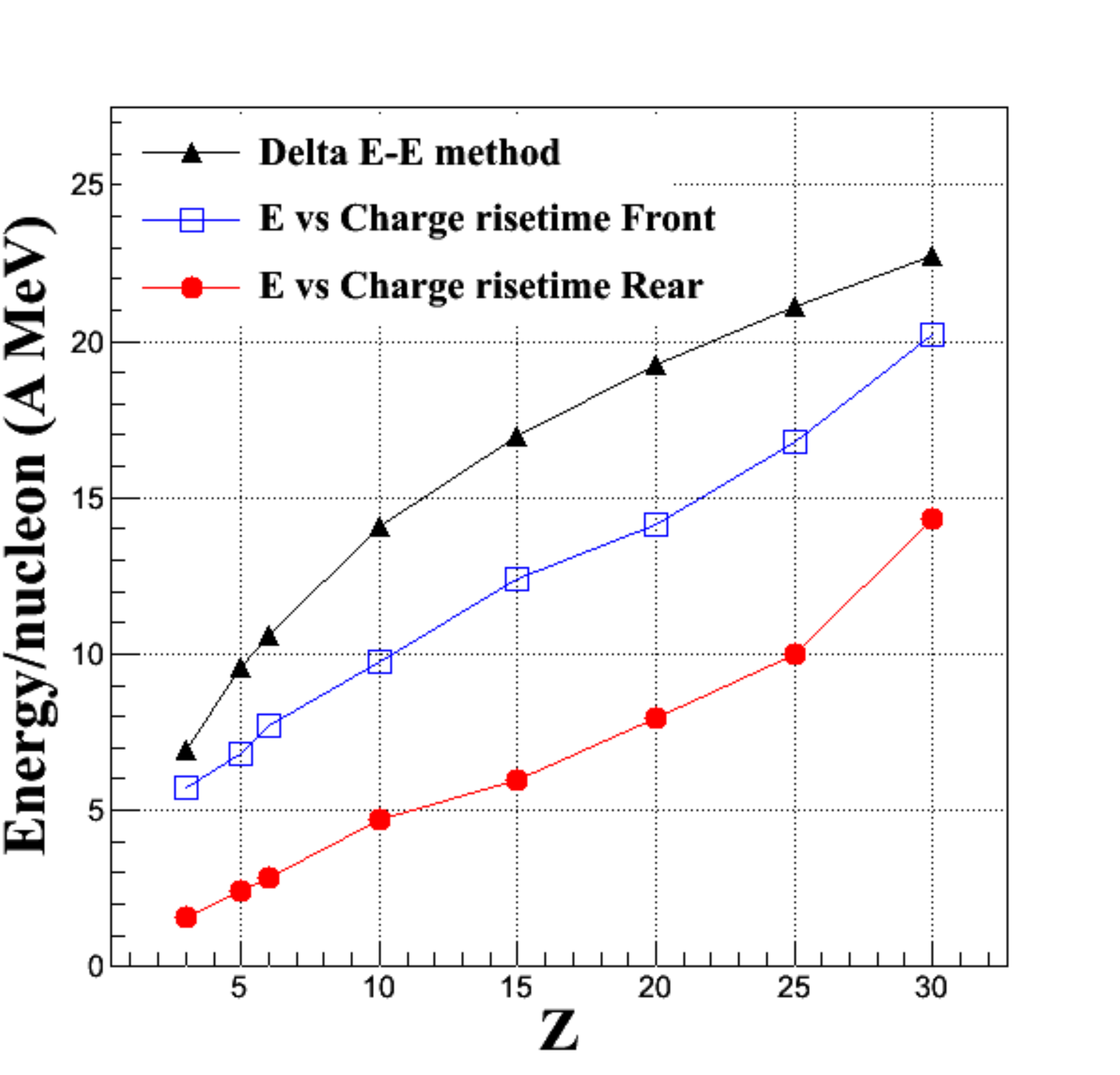}
       \caption{Energy thresholds for Z-identification.  Black
        lines and black triangles refer to the energy necessary to punch 
	through a 300~$\mu$m silicon for $\Delta$E-E identification. 
	Left part: silicon with resistivity FWHM 0.6\%, 
	using two PSA methods, low-field entrance (from~\cite{Car12}). 
	Right part: silicon with resistivity FWHM 4\%; full circles low-field 
	entrance, open squares high field entrance. (from~\cite{NLN12})} 
	\label{Fig:seuils}
   \end{minipage}%
\end{figure}

In figure~\ref{Fig:seuils} the thresholds so determined are plotted in
several cases, and compared to those imposed by the punching-through of a 
300~$\mu$m silicon when using a $\Delta$E-E method. A general remark is that
PSA does significantly lower the Z-identification thresholds. 
The left part of the figure corresponds to results obtained for the highly 
homogeneous silicon of figure~\ref{Fig:P2}, when particles enter the detector 
through the low field side. The two PSA methods, E vs $\tau _Q$
and E vs I$_{max}$, give close thresholds, from 2~\AM{} for Z=6 to 6~\AM{} 
for Z=30 and 11~\AM{} for Z=50. The right part of the figure displays data
obtained for a less homogeneous detector (4\% $\rho$-FWHM). In that case 
we compared the thresholds obtained, in the same conditions, when particles 
enter through the low or high field side~\cite{NLN12}. We firstly note 
that much lower thresholds are measured when the detector is reverse mounted, 
confirming that this is the best choice when using PSA techniques. 
Second, if we compare the thresholds obtained for rear entrance in the two 
detectors (left and right pictures), we note the influence of
the resistivity homogeneity: they are about 1.5-2.5 times lower for the
$\rho$-homogeneous detector.

\subsection{The problem of radiation damages}

As soon as silicon detectors started to be used, it was observed that the
crystal structure, and thus the detector response, was degraded by the
detected particles, particularly when very heavy ions stop in a detector. 
This raises problems when using silicon detectors with heavy beams, at angles
where they detect the high rate elastic scattering events. Any degradation
of the response of the detector becomes unacceptable when one makes use of
pulse shape analysis. A specific study of the radiation damages was
undertaken during our R\&D, and we could reach quantitave conclusions. We
irradiated a 300$\mu$m silicon detector with 35~\AM{} Xe ions. Collimators and
degraders allowed to define a zone in which the ions punch through the
detector and a second one where the ions were stopped~\cite{Bar12}.

\begin{figure}
\includegraphics[width=0.9\textwidth]{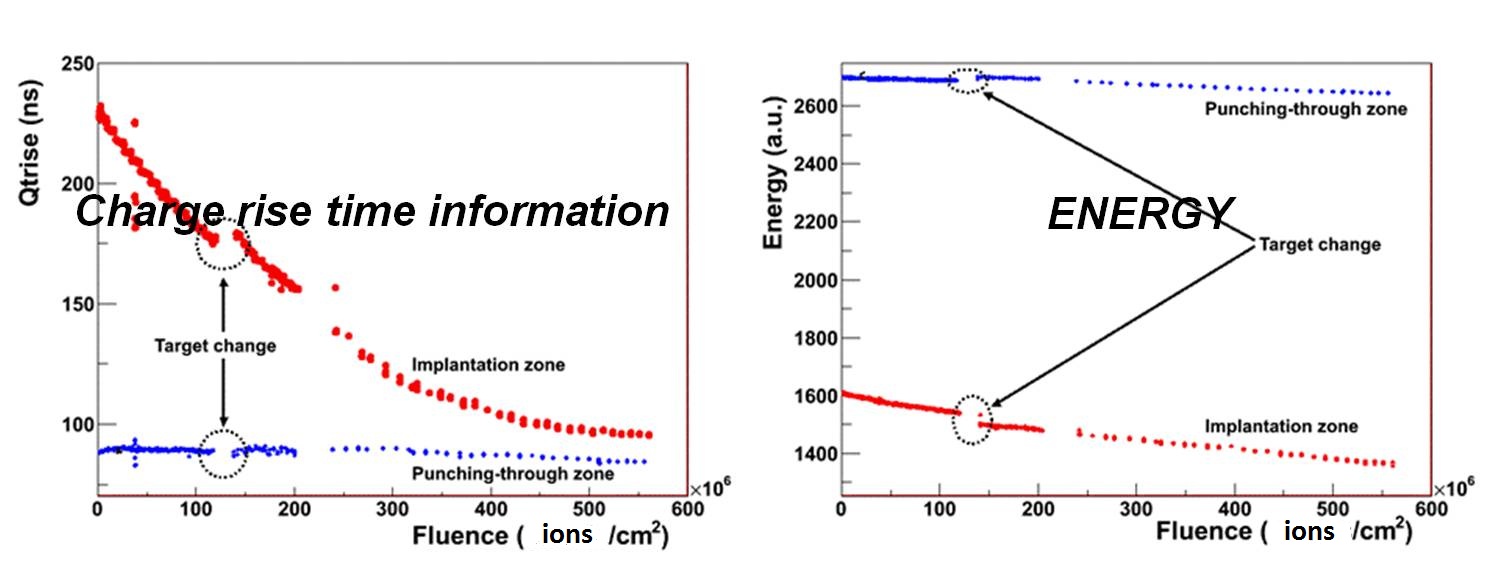}
\caption{Evolution of $\tau _Q$ and of the measured energy with the number
of detected Xe ions. In the ``implantation zone'', the ions are stopped in
the detector. (from~\cite{Bar12})} \label{Fig:RadDam}
\end{figure}

Up to 6$\times$10$^8$ ions$/$cm$^2$ ions hit the detector. Minor effects on the
measured values of E and $\tau _Q$ were observed when the Xe ions crossed the
silicon. Conversely dramatic differences were found on both parameters when
the ions stopped in the detector. At maximum fluence the energy has decreased
by 20\% whereas $\tau _Q$ dropped by a factor 2.5. 

These observations make delicate the use of very heavy ions for FAZIA
experiments, particularly at low energy where the grazing angle is large.
It would be interesting to test in the future if some annealing process
could allow to recover acceptable functioning of the damaged silicon 
detectors.

\section{FAZIA demonstrator} \label{Demo}
\begin{figure}[htb]
   \begin{minipage}[b]{0.49\textwidth}
      \includegraphics[width=\textwidth,angle=30]{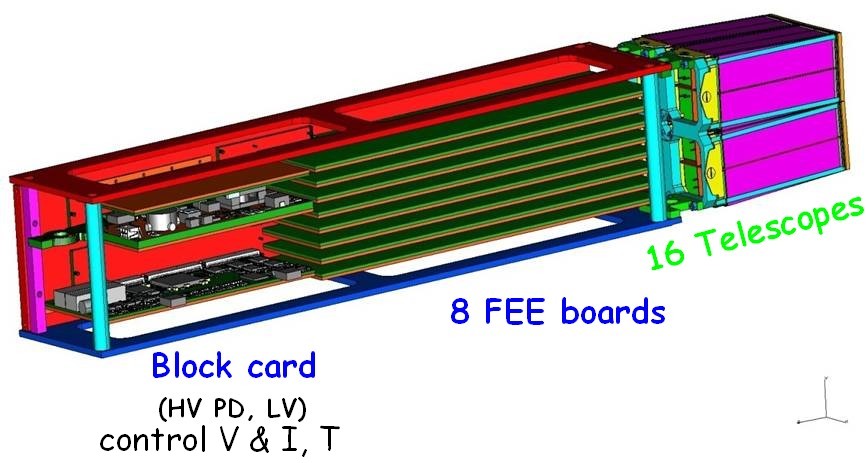}
      \caption{Mechanical arrangement of a FAZIA block.} \label{Fig:block}
   \end{minipage}%
   \hspace*{0.02\textwidth}
   \begin{minipage}[b]{0.49\textwidth}
       \includegraphics[width=\textwidth]{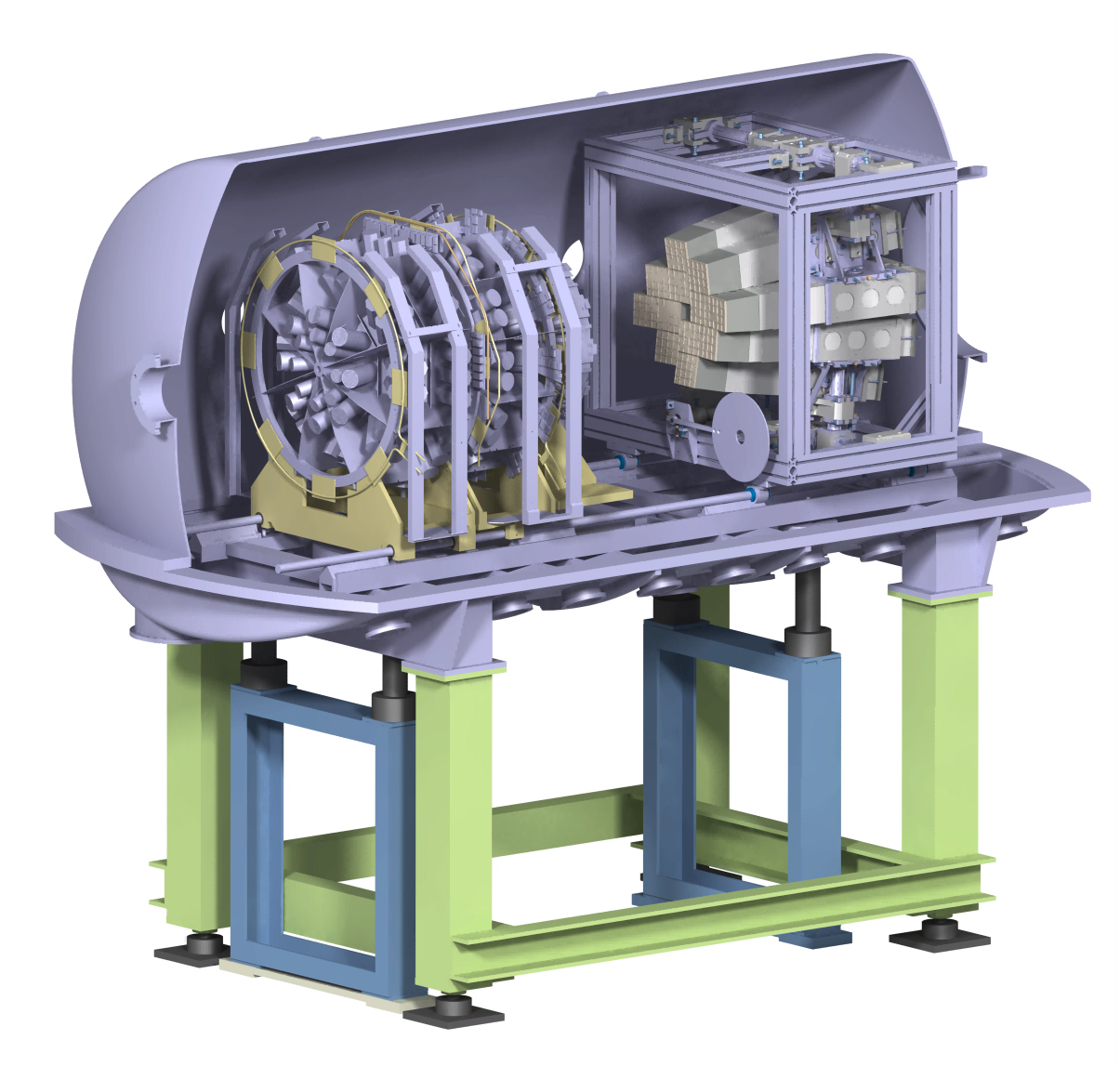}
       \caption{Mechanical coupling of the 12 blocks of the FAZIA
demonstrator (right) with rings 6-17 of INDRA inside the INDRA vacuum chamber.
(courtesy of Y. Merrer, LPC Caen).} \label{Fig:FaziaIndra}
   \end{minipage}%
\end{figure}
 The FAZIA project is presently in the phase of building a demonstrator. 
 This demonstrator will be an array of 192 telescopes arranged in 12 blocks
of 16. The final (i.e. for the future 4$\pi$ array) electronics and mechanical 
solutions are adopted. We feel useful to minimize the
detector-preamplifier-digitizer distances, therefore the FEE is located
inside the vacuum chamber\footnote{for all test results presented in the
previous section, only the preamplifiers were in vacuum, implying several
meters of differential cable between PA and FEE}. The electronic arrangement is
represented in figure~\ref{Fig:block}.
The 8 FEE boards necessary to treat the 16 telescopes are piled up and all
connected to a single block card. The block card connects the FEE to the acquisition
system by means of an optical fiber insuring the air-vacuum feed-through. 
Through this fiber it receives a 48~V voltage out of which it builds the
photodiode high voltages and different low voltages which are transmitted to
the FEE boards to generate the low voltages for preamplifiers and the high
voltages for silicon detectors.
It also houses one flash ADC to receive the accelerator RF signal which will
be used for time of flight measurement.
Due to the compacity of the electronics, an efficient cooling system is
required; all FEE and the block cards lean on a copper plate in which 
circulates a water-antifreeze liquid.

A completely equipped block will be tested at LNS (Catania) at the end of
2012. The demonstrator is expected to be ready in 2014. It will first
be coupled to existing arrays such as INDRA (GANIL-SPIRAL2), CHIMERA
(LNS-FRIBS), GARFIELD (LNL-SPES) in order to get new and improved physics
results. The modularity of the blocks allows to mount them in different
mechanical configurations, depending on the experiment. For instance dissipative
collisions will be studied by arranging the blocks in a coplanar belt at
intermediate angles. Other planes are to study fusion reactions between
light nuclei. In that aim the 12 FAZIA blocks will replace the forward rings of
INDRA, as shown in figure~\ref{Fig:FaziaIndra}. See~\cite{Cas11} for more
details on some proposed letters of intent.

\section*{References}

\providecommand{\newblock}{}



\end{document}